\documentclass[preprint,floats,aps,epsfig,nofootinbib,amssymb]{revtex4-1}
\usepackage{mathrsfs}

\usepackage{slashed}
\usepackage{graphicx,color}
\usepackage{subcaption}
\usepackage{epsfig}
\usepackage{dcolumn}
\usepackage{bm}
\usepackage[outdir=./]{epstopdf}
\newcommand{\comment}[1]{}

\begin{document}

\vspace*{-5.8ex}
\hspace*{\fill}{ACFI-T16-17}

\vspace*{+3.8ex}

\title{Tau flavored dark matter and its impact on  tau Yukawa coupling}

\author{Wei Chao$^{1}$}
\email{chao@physics.umass.edu}

\author{Huai-Ke Guo$^{1}$} 
\email{huaike@physics.umass.edu}

\author{Hao-Lin Li$^{1}$}
\email{haolinli@physics.umass.edu}

 \affiliation{Amherst Center for Fundamental Interactions, Department of Physics, University of Massachusetts-Amherst
Amherst, MA 01003 USA 
 }

\vspace{3cm}

\begin{abstract}

In this paper we preform  a systematic study of the tau flavored dark matter model by introducing two kinds of mediators (a scalar doublet and a charged scalar singlet). The electromagnetic properties of the dark matter, as well as their  implications in dark matter direct detections, are analyzed in detail. The model turns out contributing a significant radiative correction to the tau lepton mass,  in addition to loosing the tension between the measured dark matter relic density and constraints of dark matter direct detections. The loop corrections can be ${\cal O}(10\%)$ of the total tau mass. Signal rates of the Higgs measurements  from the LHC in the $h\to\tau \tau$ and $h\to \gamma \gamma$ channels, relative to the Standard Model expectations, can be explained in this model.

\end{abstract}

\maketitle
\section{Introduction}

Accumulated cosmological and astrophysical observations have confirmed the existence of the cold dark matter 
$\Omega_{} h^2 =0.1199\pm0.0022$~\cite{Planck:2015xua},
which requires an extension to the minimal Standard  Model (SM). 
Since the nature of the dark matter and the way it interacts with the SM particles remain mysteries, 
it catalyzes various portals of the dark matter. Strengths of interactions between ordinary matter  and dark matter are 
severely constrained by the observed dark matter relic abundance and exclusion limits from dark matter direct 
and indirect detections. 
Thanks to the advancement of the technology, many well-motivated dark matter models were  tested and
excluded by the dark matter direct and(or) indirect detection experiments! It raises challenge to the dark matter 
model building, but shows people hope of discovering dark matter in the laboratory. 

Among various dark matter models, flavored dark matter~\cite{Chang:2014tea,Cai:2014hka,Baltz:2002we,Chao:2010mp,Schmidt:2012yg,Agrawal:2011ze,Cohen:2009fz,Bai:2014osa,Freitas:2014jla,Bai:2013iqa,DiFranzo:2013vra,Cheung:2013dua,Agrawal:2015tfa,Calibbi:2015sfa,Kilic:2015vka,Lee:2014rba,Agrawal:2014aoa,Agrawal:2014una,Batell:2013zwa,Kile:2013ola,Kumar:2013hfa,Kile:2011mn,Yu:2014mfa,Chen:2015jkt} is  interesting and appealing for the following three reasons: (1) It may naturally explain the galactic center  gamma ray 
excess~\cite{Agrawal:2014una} observed by the {\it Fermi-LAT}~\cite{FermiLAT:2011ab}. (2) It is tightly 
connected with the flavor physics. (3) It may release the tension between 
the observed dark matter relic density and constraints from underground laboratory direct detections.  
Besides, collider searches of the flavored dark matters are accessible and it was pointed out that collider searches are 
remarkably complementary for the quark(lepton)-portal dark matter models~\cite{Chang:2014tea,Bai:2013iqa,Bai:2014osa,Kile:2011mn,Yu:2014mfa}. 
Notice that dark matter could be incorporated into numerous models of flavors. It deserves a systematic study
of these models. 

In this paper we study phenomenologies of the tau lepton flavored dark matter by assuming dark matter is a 
Dirac fermion and mainly couples to the third generation leptons with an extra scalar doublet and a charged scalar 
singlet as mediators. We focus on the following aspects of this model: i) the dark matter electromagnetic form factors, ii)its relic density, iii) signatures in the dark 
matter direct detection and iv) the impact of the model to the $h\bar\tau\tau$ coupling as well as $h\gamma \gamma$ 
coupling, where $h$ is the SM-like Higgs. Our findings can be summarized as the followings:
\begin{itemize}
\item 

The tension between the dark matter relic density and direct detection is highly loosed even for  ${\cal O} (1)$ 
 Yukawa couplings of the dark matter with tau lepton. The charge radius of the dark matter dominates the 
 scattering of the dark matter with the nuclei for light dark matter case while the magnetic moment plays more 
 important role  in direct detections for the heavy dark matter case.

\item The Yukawa coupling of $ h\bar \tau \tau$  can be significantly  changed  compared to the 
  SM case. Tau lepton mass arises from two parts in this model: the general Yukawa interaction and the one-loop radiative correction, whose effect 
  is roughly proportional to the dark matter mass. It turns out the loop effect can contribute  about ${\cal O}(10\%)$ 
  of the total tau mass.

\item Higgs to diphoton decay rate can be slightly modified in this model. The ratio $\mu_{\gamma \gamma}$ can 
  be in the range $(0.7,~1.25)$, which is still consistent the current bounds given by the  ATLAS and CMS collaborations.
\end{itemize}
The signatures of the dark matter at colliders are also briefly discussed in the paper.  Compared with previous studies of lepton portal dark matter models , as was mentioned in the references, our studies are new in the 
following aspects: (1) We focus on the multi-mediator scenario; (2) The Yukawa couplings of $h\bar \tau \tau$ and $h\gamma \gamma$  can be significantly modified in this model.   

Let's comment on a possible extension of this tau flavored dark matter model. If one assumes the dark matter 
couples both to the muon and tau leptons, then the observed Higgs to $\tau \mu$ decay rate~\cite{Khachatryan:2015kon,Aad:2015gha}
can be generated, but it also gives an overly estimated branching ratio of $\tau \to \mu \gamma$.  
Thus the lepton flavored  dark matter model can hardly explain the Higgs lepton-flavor-violating decays.

The remaining of this paper is organized as follows:  We briefly describe our model in section~\ref{sec:model}. Section~\ref{sec:pheno} is focused on the 
phenomenologies of the model, including dark matter relic density,  signatures in direct detections, the $\tau$-lepton mass, $h \to\bar \tau \tau $ and $h\to \gamma \gamma $ decay rate. The last part is concluding remarks.

\section{\label{sec:model}Model}

We extend the SM with an inert scalar doublet, a singly charged scalar singlet and a Dirac dark matter, which is 
stabilized by a $Z_2$ discrete flavor symmetry, under which dark matter and the third generation leptons are odd while
all other particles are even. In the following we first describe scalar interactions, then go to the dark matter interactions. 
The scalar potential can be written as
\begin{eqnarray}
V&=& -\mu_{}^2 H^{\dagger} H +  \lambda (H^{\dagger} H)^2 + m_{1}^2 \Phi^\dagger \Phi 
+ \lambda_1(\Phi^\dagger \Phi)^2 + \lambda_2(\Phi^\dagger \Phi)( H^\dagger H )  +\lambda_3  (\Phi^\dagger H)( H^\dagger \Phi)  
\nonumber \\&& + m_2^2 S^+S^- + \lambda_4 (S^+ S^-)^2
+ \lambda_5 (S^+S^-) (H^\dagger H)+ \lambda_6  (S^+S^-) (\Phi^\dagger \Phi) \nonumber \\
&&+ \sqrt{2}\Lambda H^T  \varepsilon \Phi S ^-+ {\rm h.c.} \; , \label{potential}
\end{eqnarray}
where $H^T\equiv(G^+, (h + i G_0+ v)/\sqrt{2})$ is the SM Higgs doublet, $v=246~{\rm GeV}$ is the vacuum expectation value  (VEV), 
$\Phi^T\equiv(\Phi^+, (\rho + i \eta)/\sqrt{2})$ is the inert scalar doublet, $S^\pm$ is the singly charged scalar 
singlet and $\Lambda$ has mass dimension +1.  Assuming that the mass term of $\Phi$ is positive,  it develops no 
VEV. As a result, there is no mixing between $h$ and $\rho$. The masses of neutral 
scalars can be written as
\begin{eqnarray}
  m_h^2 = 2 \lambda v^2 \; , \hspace{1cm} m_\rho^2 = m_\eta^2 = m_1^2 +{1\over 2 }(\lambda_2 + \lambda_3 ) v^2 \; .
\end{eqnarray}
Due to the last term in Eq. (\ref{potential}), there is mixing between $\Phi^+$ and $S^+$ and the relevant mass matrix is
\begin{eqnarray}
\left( \matrix{ m_1^2 + {1\over 2 } \lambda_2  v^2 
 &  -\Lambda v 
\cr
-\Lambda v 
& m_2^2 + {1\over 2 } \lambda_5 v^2 
} 
\right) .\label{massmatrix}
\end{eqnarray}
The corresponding mass eigenvalues are
\begin{eqnarray}
\hat{m}^2_{1,2} = {1 \over 2 } \left\{  m_1^2 + m_2^2 + {1 \over 2 } (\lambda_2 + \lambda_5) v^2 \pm \sqrt{ \left[m_1^2 -m_2^2 + {1\over 2 } (\lambda_2 -\lambda_5) v^2 \right]^2 + 4 (\Lambda v)^2} \right\} ,
\end{eqnarray}
and the relations between physical eigenstates and interaction eigenstates are $\Phi^+ = c_\theta \hat{\Phi}^+ + s_\theta \hat{S}^+$, 
$S^+=-s_\theta \hat{\Phi}^+ + c_\theta \hat{S}^+$, where $c_\theta= \cos \theta$ and $s_\theta=\sin \theta$, with $\theta$ the 
rotation angle that diagonalizes the mass matrix in Eq.\ref{massmatrix}.

Free parameters of this model are: $m_h,~m_\rho, ~\hat{m}_{1,2}, ~\theta,~\lambda_i(i=1,2,4,5,6)$, 
and $\Lambda$.  Unphysical parameters in the Higgs potential can be written in terms of the physical parameters: 
\begin{eqnarray}
(A)\left\{
\begin{array}{l}
\mu^2=1/2 m_h^2\\
m_1^2 = \hat{m}_1^2 c_\theta^2 + \hat{m}_2^2 s_\theta^2-1/2 \lambda_2 v^2 \\
m_2^2 = \hat{m}_1^2 s_\theta^2 + \hat{m}_2^2 c_\theta^2-1/2 \lambda_5 v^2
\end{array}
\right.
\hspace{1cm}
(B)\left\{
\begin{array}{l}
\lambda_{ ~}=m_h^2 v^{-2}/2\\
\Lambda= (\hat{m}_1^2-\hat{m}_2^2) c_\theta s_\theta v^{-1} \\
\lambda_3 = 2v^{-2}[m_\rho^2-(\hat{m}_1^2 c_\theta^2 + \hat{m}_2^2 s_\theta^2)]
\end{array}
\right.
\label{parameters}
\end{eqnarray}
Notice that in the parameter set we have chosen, $\lambda_{1}, \lambda_{4}, \lambda_6$ describe quartic interactions 
among these extra scalars and are not so relevant for the study in this paper. $\lambda_2$ and 
$\lambda_5 $ are relevant for the $h\gamma \gamma$ and $h\bar\tau \tau$ couplings as will be seen in the next section.

%
\begin{figure}
\centering
  \begin{subfigure}[b]{0.45\textwidth}
  \includegraphics[width=\textwidth]{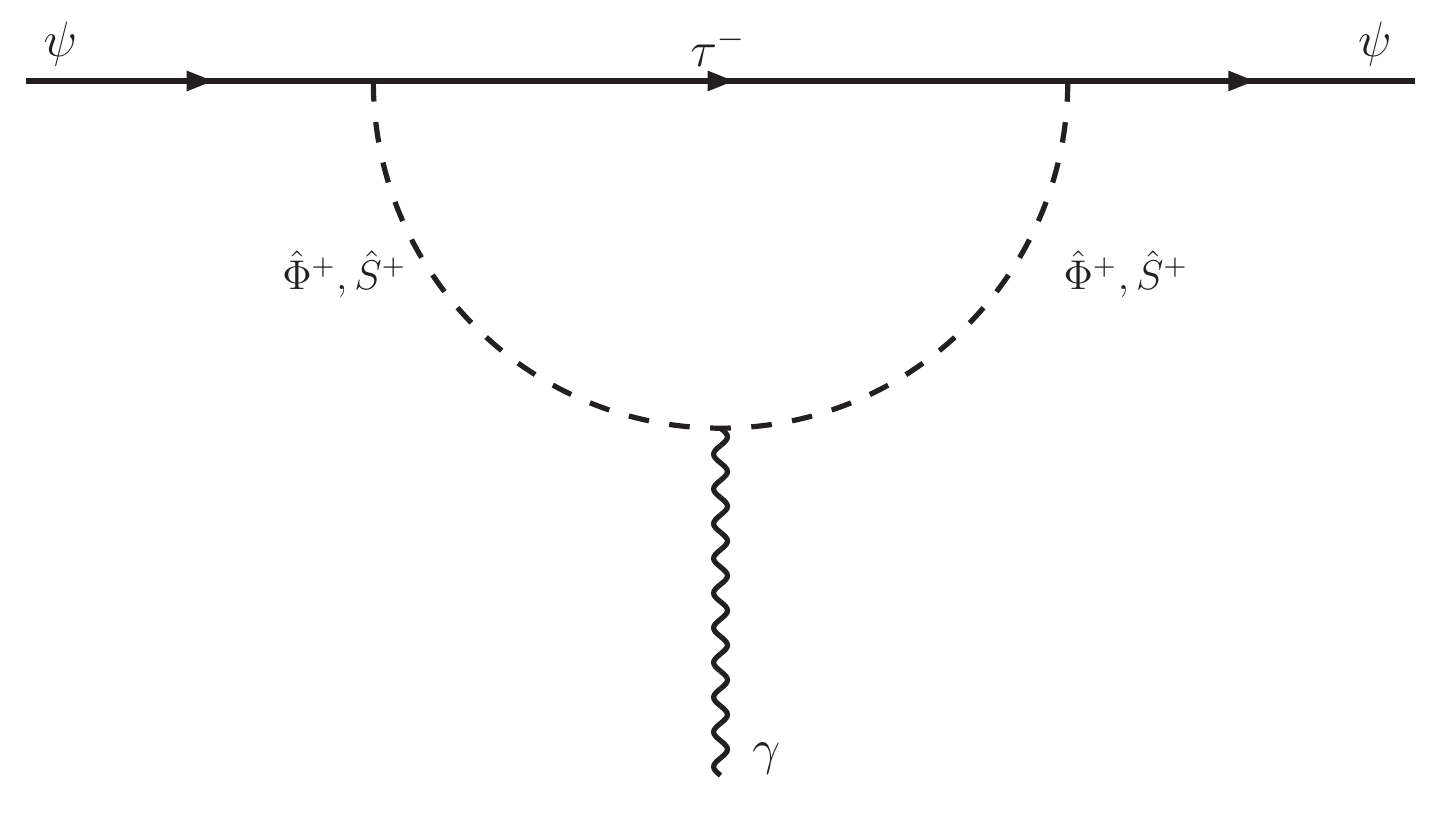}
  \end{subfigure}
  \begin{subfigure}[b]{0.45\textwidth}
  \includegraphics[width=\textwidth]{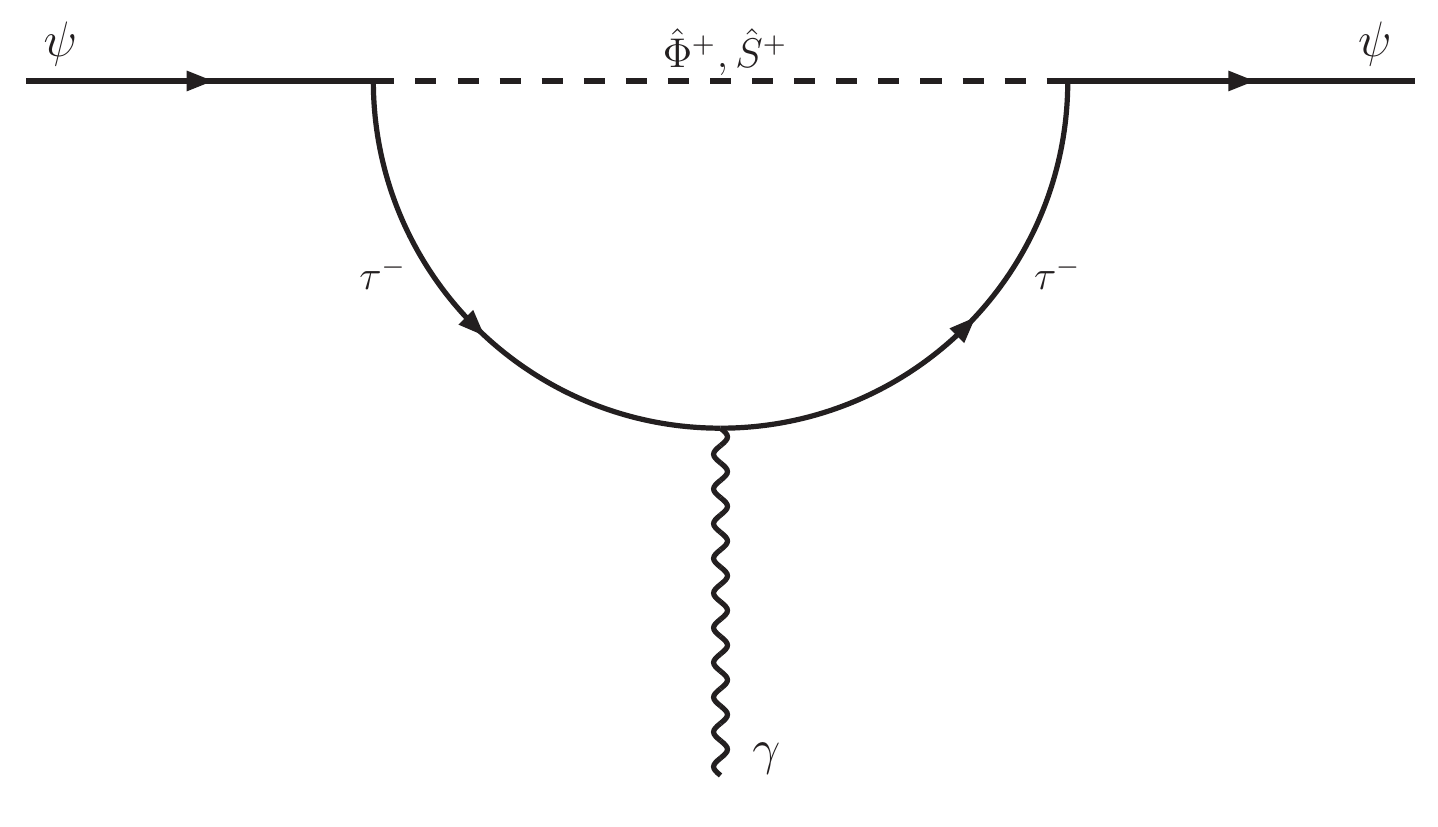}
  \end{subfigure}
\caption{\label{ddloopgamma} Feynman diagrams contributing to the dark matter electromagnetic form factors.}
\end{figure}
%

We assume that dark matter only interacts with the new scalars and third generation leptons which can be written as
\begin{eqnarray}
  \label{portalL}
-{\cal L}_Y = \kappa_1^{}  \overline {\ell^3_L} \tilde \Phi \psi + \kappa_2^{} \overline \psi S^+ \tau_R + {\rm h.c.}  \; ,
\end{eqnarray}
where $\ell_L^3$ is the third generation left-handed lepton doublet, $\tau_R$ is the right-handed tau lepton and $\psi$ is the Dirac dark matter. As a result, 
the dark matter can only annihilate into $\bar{\tau}\tau$ and $\bar{\nu}_\tau {\nu}_\tau$. For the benefit of 
the direct detection, one needs to calculate the electromagnetic form factors of the dark matter, which arise at 
one loop level from the relevant Feynmann diagrams shown in Fig.~\ref{ddloopgamma}. The induced effective dark
matter-photon interactions are 
\begin{eqnarray}
\Delta \mathcal{L_{\text{EM}}^{\psi}} =  b_{\psi} \bar{\psi} \gamma^{\mu} \psi \partial^{\nu} F_{\mu\nu} 
+ c_{\psi} \bar{\psi} \gamma^{\mu} \gamma^5 \psi \partial^{\nu} F_{\mu\nu} +
\frac{\mu_{\psi}}{2} \bar{\psi} \sigma^{\mu\nu} \psi F_{\mu\nu} ,
  \label{}
\end{eqnarray}
where $b_\psi$ is the charge radius, $c_\psi$ is the axial charge radius or anapole moment and $\mu_\psi$ is the 
magnetic moment. Since there is no CP violation in the dark matter sector, the electric dipole moment term is absent.  We assume the 
following mass hierarchy $m_{\tau}\ll m_{\psi}< \hat{m}_{1,2}, m_{\rho,\eta}$.  Besides the typical momentum transfer of DM-Nucleon interactions is  about  $50~\text{MeV}$, thus the momentum transfer, $\sqrt{-q^2}$, is far smaller 
than the $\tau$ mass  and constitutes the smallest scale.  Collecting all the contributing diagrams and expanding 
in terms of $q^2$, we obtain
\begin{eqnarray}
  \mu_{\psi} &=& \sum_{i = 1}^2-\frac{e m_{\psi}  \zeta_i }{64\pi^2} \int^1_0 dx \frac{x(1-x) }{\Delta_i} \; , \nonumber \\
  b_{\psi} &=&\sum_{i = 1}^2{e \zeta_i \over 32 \pi^2 } \int_0^1 d x \left\{{ x^3 -2(1-x)^3\over 6 \Delta_i } + {(x-1)^3 (x^2 m_\psi^2+m_\tau^2)  +2(1-x)x^4 m_\psi^2 \over 6 \Delta_i^2 }  \right\} , \label{electromag}\\
   c_{\psi} & =&  \sum_{i = 1}^2 \frac{e \hat{\zeta}_i}{192 \pi^2} \int^1_0  dx 
   \left\{
    \frac{ (-3x^3+6x^2-6x+2) x \hat m_i^2+(-2x^4+6x^3-9x^2+7x-2) x m_{\psi }^2}{\Delta_i^2} \right\} ,\nonumber
\end{eqnarray}
where $m_\psi$ is the dark matter mass, 
$\zeta_1=c_\theta^2 \kappa_1^2+s_\theta^2 \kappa_2^2$, 
$\zeta_2=s_\theta^2 \kappa_1^2+c_\theta^2 \kappa_2^2$, 
$\hat{\zeta}_1 = c_{\theta}^2 \kappa_1^2  -  s_{\theta}^2 \kappa_2^2$, 
$\hat{\zeta}_2 = s_{\theta}^2 \kappa_1^2  -  c_{\theta}^2 \kappa_2^2$,
and $\Delta_i= x \hat{m}_i^2 + x(x-1)m_\psi^2 + (1-x)m_\tau^2$.  We have ignored terms proportional to 
${\cal O} (m_\tau^2)$ in Eg. (\ref{electromag}). Note that the limit $m_{\psi}, m_{\tau} \ll \hat{m}_{1,2}$ allows us to recover 
the familiar result~\cite{Chang:2014tea,Hamze:2014wca}
\begin{eqnarray}
  b_{\psi} = \sum_{i}\frac{e \zeta_i^2}{64\pi^2 \hat{m}_{i}^2} \left(1+\frac{2}{3} \ln \frac{m_{\tau}^2}{\hat{m}_{i}^2}\right) ,
  \label{}
\end{eqnarray}
where $m_{\tau}$ serves as an infrared regulator.

Similarly there are also form factors for the effective dark matter-Z boson interactions. The contribution of 
these interactions to the dark matter-nuclei scattering cross section is subdominant compared with those arising 
from electromagnetic form factors. So we neglect these interactions in our calculation. 

\section{\label{sec:pheno}Phenomenology}

We will study in this section phenomenologies arising from this model, including the dark matter relic density, 
signatures in direct detections,  the loop induced $\tau$ lepton mass, the effective coupling of $h\bar \tau \tau$ as well as 
the Higgs to diphoton decay rate.  Finally we will discuss signatures of our model at colliders.

\subsection{Relic density}

\begin{figure}
\centering
   \begin{subfigure}[b]{0.45\textwidth}
   \includegraphics[width=\textwidth]{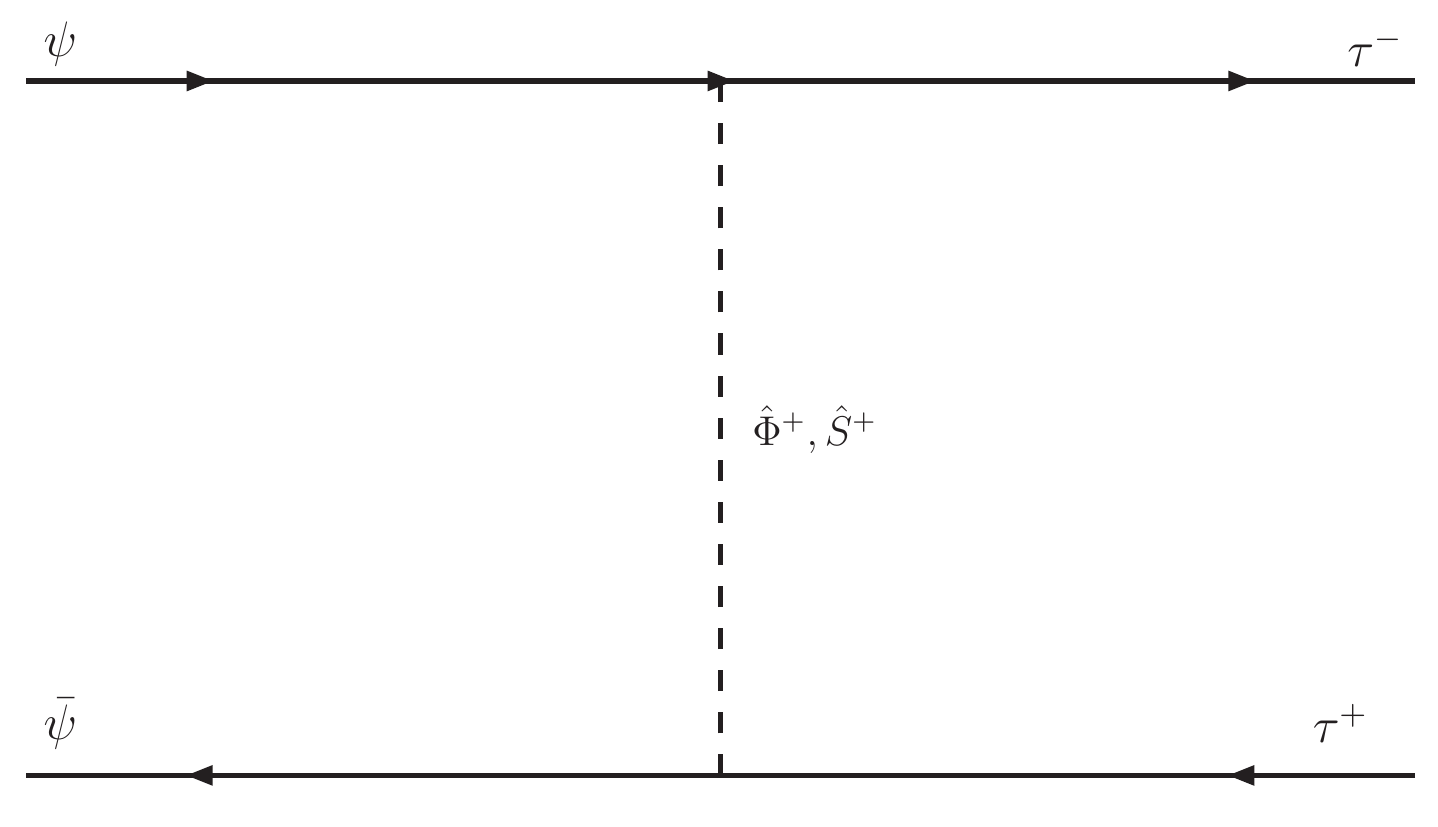}
   \end{subfigure}
   \begin{subfigure}[b]{0.45\textwidth}
   \includegraphics[width=\textwidth]{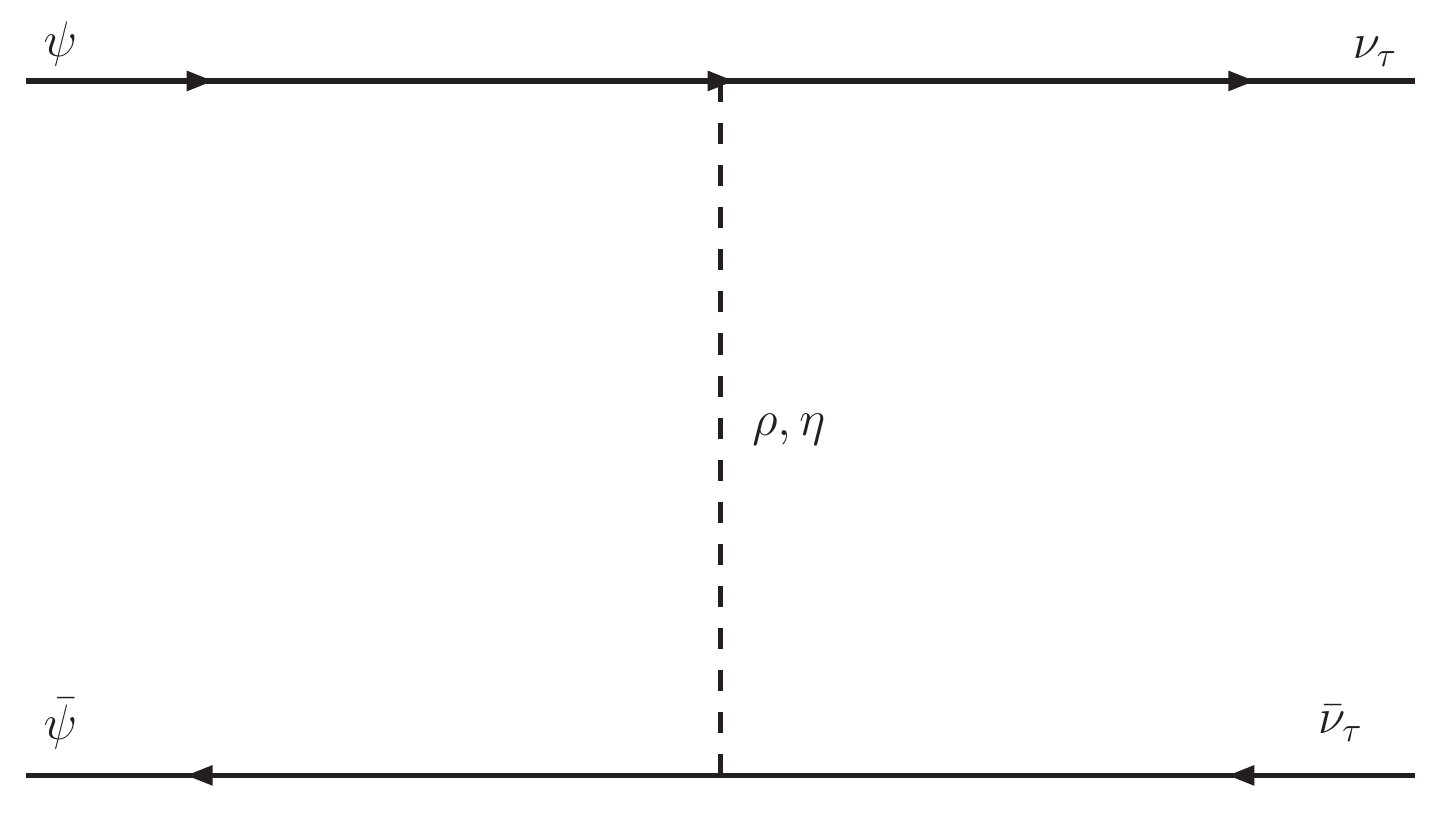}
   \end{subfigure}
 \caption{\label{fig:relic} Dark matter annihilation channels.}
 \end{figure}
 
We have assumed that the dark matter is a Dirac fermion and only interacts with the third generation leptons in our model. 
It annihilates into $\bar{\tau} \tau/\bar{\nu}_\tau \nu_\tau$ with the relevant Feynman diagrams shown in Fig.~\ref{fig:relic}.
The cold dark matter was in local thermodynamic equilibrium  in the early Universe. When its interaction rate drops below 
the expansion rate of the Universe, the dark matter is said to be decoupled.  The evolution of the dark matter number density $n$,
is governed by the Boltzmann equation~{\cite{Gondolo:1990dk}}:
\begin{eqnarray}
\dot{n} + 3 Hn =- \langle \sigma v_{\rm M\slashed{o}ller} \rangle ( n^2 -n_{\rm EQ}^2 ) \; ,
\end{eqnarray}  
where $H $ is the Hubble constant, $\sigma v_{\rm M\slashed{o}ller}$ is the total annihilation cross section multiplied by  
the M$\slashed{\rm o}$ller velocity with $v_{\rm M\slashed{o}ller}=(|v_1 -v_2 |^2 -|v_1 \times v_2 |^2 )^{1/2}$, brackets denote 
thermal average and $n_{\rm EQ}$ is the number density in thermal equilibrium. It has been shown that 
$\langle \sigma v_{\rm M\slashed{o}ller} \rangle =\langle \sigma v_{\rm lab} \rangle = 1/2 [1 + K_1^2 (x) /K_2^2 (x)] \langle \sigma v_{\rm cm} \rangle$~{\cite{Gondolo:1990dk}}, 
where $x=m_{\rm DM}/T$ and $K_i (x)$ is the modified Bessel functions of the  $i$-th order. To derive the relic density of the tau flavored dark matter, 
one needs to calculate the thermal average of the total annihilation cross section. Analytically one can approximate the thermal average 
$\langle\sigma v\rangle$ with the non-relativistic expansion $\langle \sigma v \rangle =a + b \langle v^2 \rangle $ in the lab frame,
\begin{eqnarray}
\langle \sigma v\rangle &=& \sum_{i=1}^4 \zeta_i^2  \left( {m_\psi^2 \over 32 \pi (m_\psi^2 + \hat{m}_{i}^2)^2 } + \langle v^2 \rangle {m_\psi^2 (-7 m_\psi^4 -18 m_\psi^2 \hat{m}_i^2 +\hat{m}_i^4 )\over 384 \pi (m_\psi^2 + \hat{m}_i^2)^4}\right)+ \nonumber \\
&& {1\over 4} s_{2\theta}^2 (\kappa_1^2-\kappa_2^2)^2\left( { m_\psi^2 \over 16 \pi (\hat{m}_1^2 + m_\psi^2 ) (\hat{m}_2^2 + m_\psi^2 ) } + { \langle v^2 \rangle  \Delta \over 192 \pi  (\hat{m}_1^2 + m_\psi^2 )^3 (\hat{m}_2^2 + m_\psi^2 )^3 }\right) \nonumber \\
&\equiv& a + b \langle v^2 \rangle , \label{thermalaverage}
\end{eqnarray}
where
\begin{eqnarray}
\Delta&=& -m_\psi^2 \left( 7 m_\psi^8+16m_\psi^6(\hat{m}_1^2+\hat{m}_2^2)+m_\psi^4 (5\hat{m}_1^4+32 \hat{m}_1^2 \hat{m}_2^2 + 5 \hat{m}_2^2)\right.\nonumber \\&&+ 8 m_\psi^2 \hat{m}_1^2 \hat{m}_2^2 (\hat{m}_1^2 + \hat{m}_2^2) \left.-\hat{m}_1^4 \hat{m}_2^4\right) .
\end{eqnarray}
Here $\zeta_{1,2}$ were defined below Eq. (\ref{electromag}) and $\zeta_{3,4}=\sqrt{2}\kappa_1^2$. The notation $\hat{m}_i$, where $i=1,2,3,4$, 
denotes the mass of $\hat{\Phi}^+$, $\hat{S}^+$, $\rho$ and $\eta$ respectively.

The present relic density of the DM is simply given by 
$\rho_{\rm DM } = m_{\rm DM}  n_{\rm DM} = m_{\rm DM} s_0 Y_{\infty} $~\cite{Bertone:2004pz},  where $s_0$ is the present 
entropy density. The relic abundance can be written in terms of the critical density
\begin{eqnarray}
  \Omega h^2 \approx 2 \times{1.07\times 10^9 \over M_{\text{pl}}} {x_F \over \sqrt{g_*} } {1 \over a + 3 b/x_F} ,
\end{eqnarray}
where $a$ and $b$ were defined in  Eq. (\ref{thermalaverage}), $M_{\rm  pl}$ is the Planck mass, $x_F=m_{\rm DM}/T_F$ with 
$T_F$ being the freezing  out temperature of the dark matter, $g_*$ is the degrees of freedom at the freeze out temperature 
and the factor $2$ on the right-hand side accounts for the fact that dark matter in our model is a Dirac fermion.

\begin{figure}
\centering
\scalebox{0.95}{
\begin{subfigure}[b]{0.49\textwidth}
  \includegraphics[width=\textwidth]{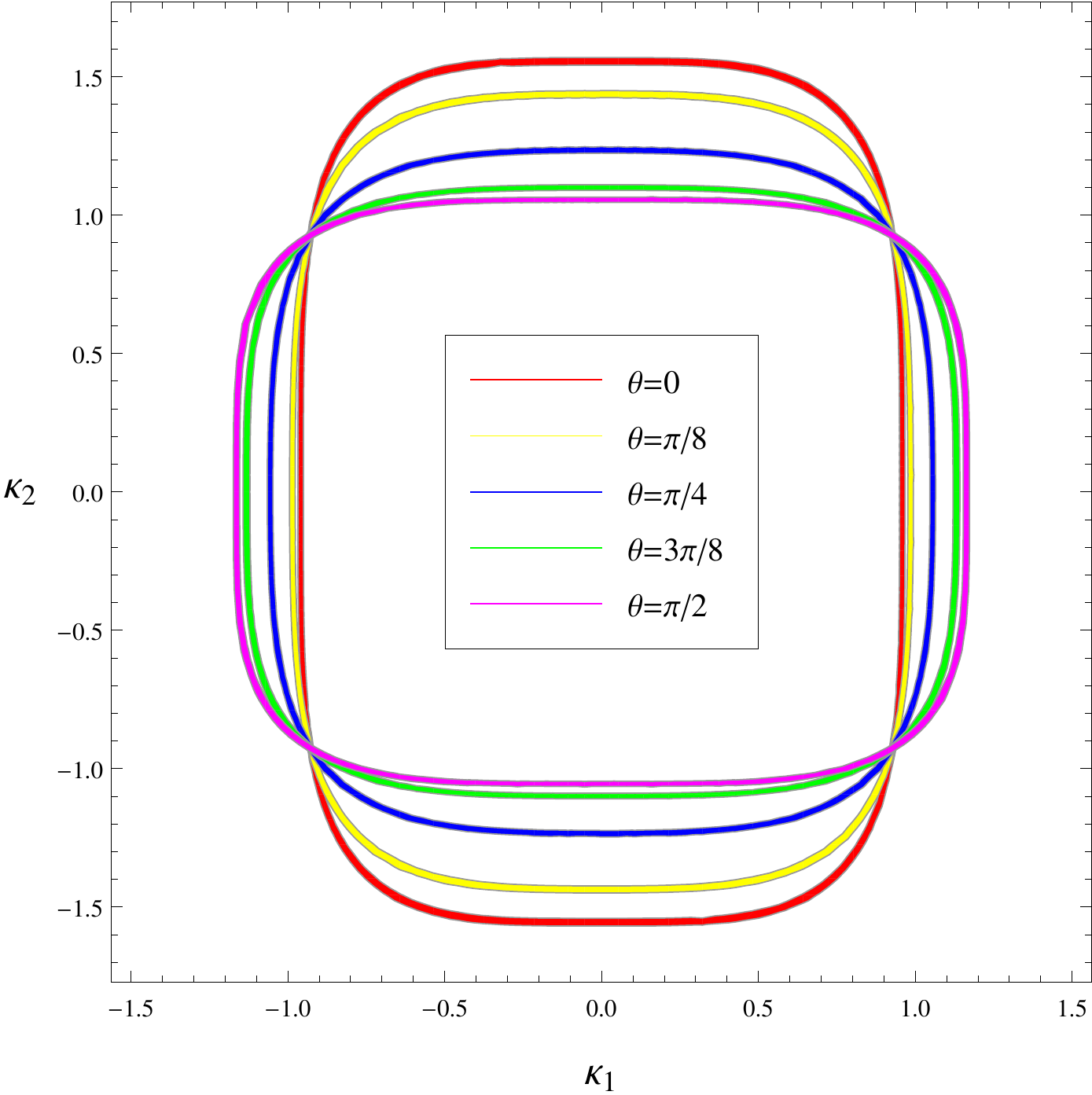}
  \caption{}
\end{subfigure}
}
\scalebox{0.95}{
\begin{subfigure}[b]{0.49\textwidth}
  \includegraphics[width=\textwidth]{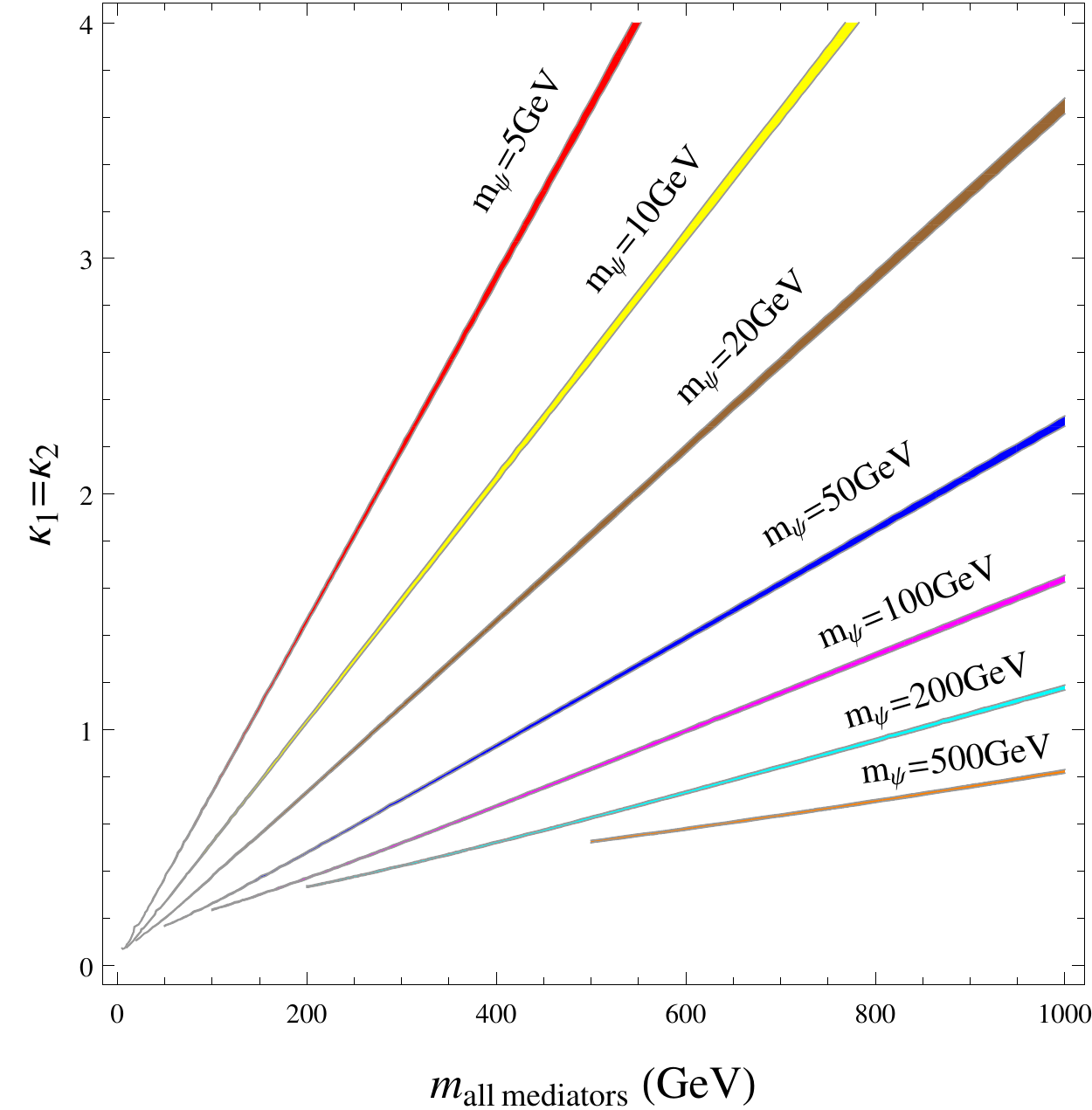}
  \caption{}
\end{subfigure}
}
\\
\scalebox{0.95}{
\begin{subfigure}[b]{0.49\textwidth}
  \includegraphics[width=\textwidth]{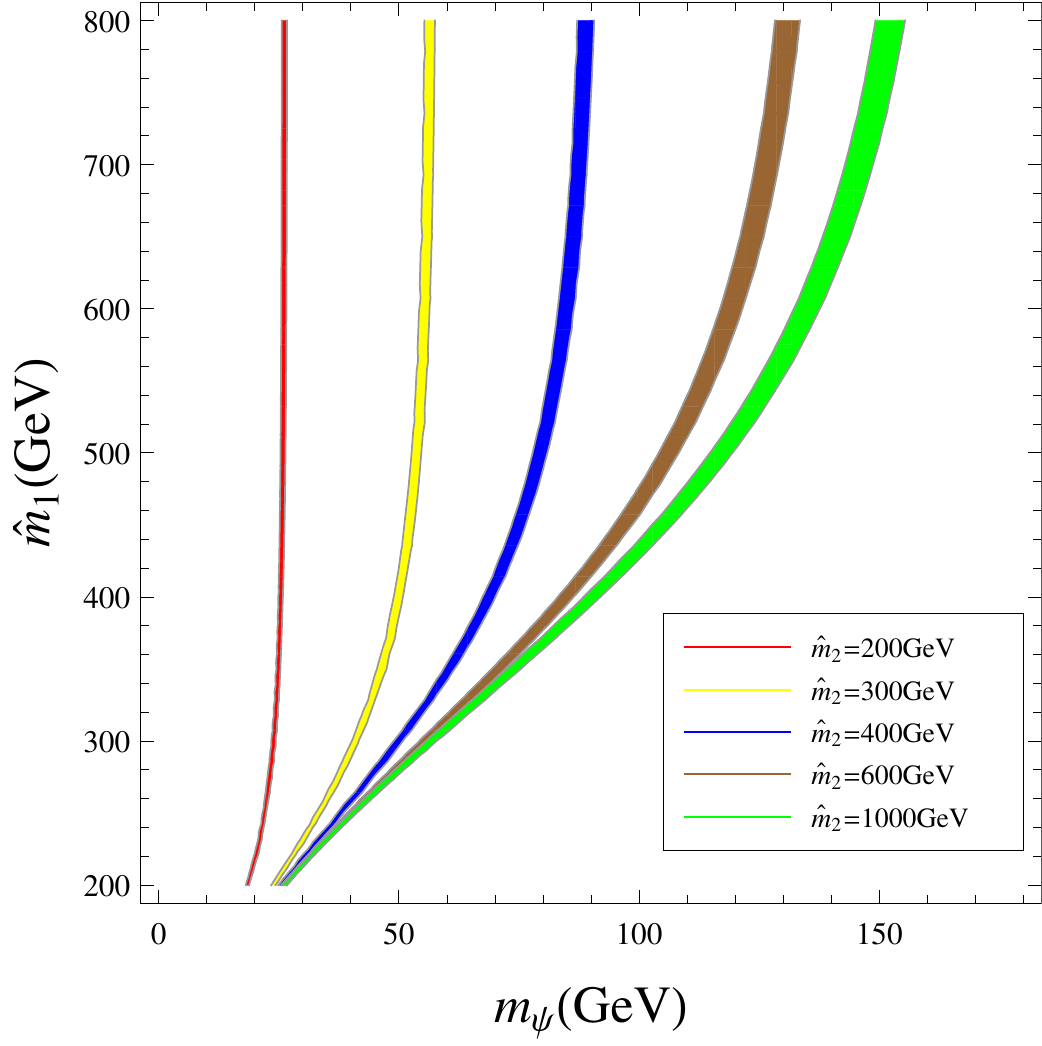}
  \caption{}
\end{subfigure}
}
\scalebox{0.95}{
\begin{subfigure}[b]{0.49\textwidth}
  \includegraphics[width=\textwidth]{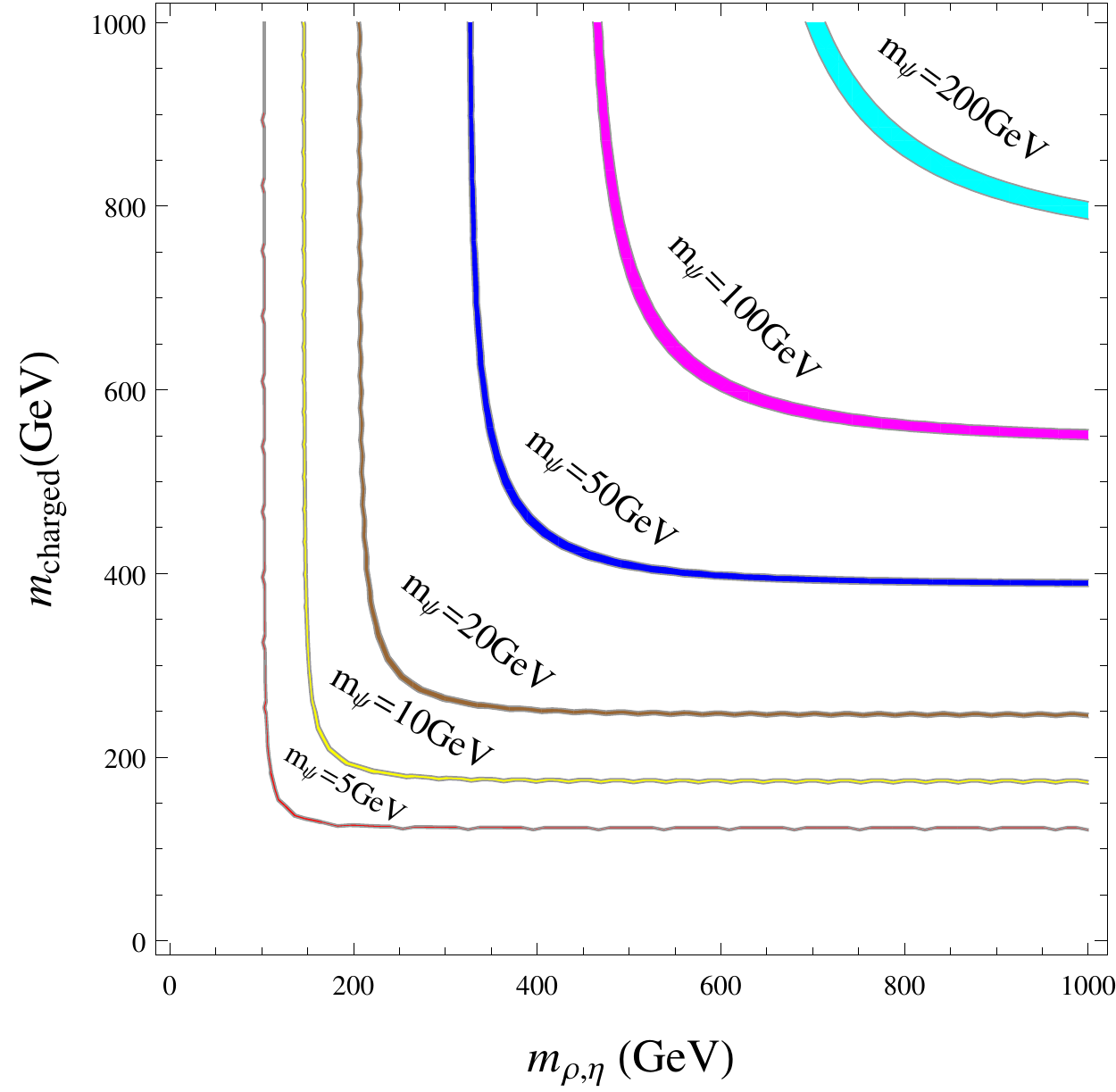}
  \caption{}
\end{subfigure}
}
\caption{\label{relicPlots}
The contours of the relic density within two standard deviations of the measured value: In Fig. (a), we show contours in
$\kappa_1-\kappa_2$ plane with different inputs of mixing angle $\theta$,  by setting $m_{\psi}=100$ $\text{GeV}$,
$\hat{m}_1=400$ $\text{GeV}$, $\hat{m}_2 =600$ $\text{GeV}$ and $m_{\rho}=m_{\eta}=700$ $\text{GeV}$; In Fig. (b), we show contours in the $\kappa_{1}-m_{\rm all ~mediators}$ plane for different dark matter masses, by assuming all mediators have the same mass and  $\kappa_1=\kappa_2$; Fig.(c) show contours in the $\hat{m}_{1}$ - $m_{\psi}$ plane for different values of 
$\hat{m}_{2}$, by setting $\kappa_1=\kappa_2=1$ and $m_{\rho}=m_{\eta}=700$ $\text{GeV}$;  In Fig. (d), we set $\kappa_1 = \kappa_2 = 1$ and plot charged mediator versus neutral mediator masses for several dark matter masses.
}
\end{figure}

The dark matter relic density measured by the Planck experiment is $\Omega h^2 =  0.1199\pm0.0022$~\cite{Planck:2015xua}.
To check its constraint on the parameter space, we plot in Fig. \ref{relicPlots} (a)  contours of the dark matter relic 
density requiring the relic density to be within two standard deviations of the measured central value in the 
$\kappa_1^{}-\kappa_2^{}$ plane by setting  $\hat m_{1}=400~\text{GeV}$, $\hat m_{2}=600~\text{GeV}$ and 
$m_{\rho}=m_{\eta}=700~\text{GeV}$. The  red, yellow, blue, green and pink contours correspond to 
$\theta = 0, ~\pi/8,~\pi/4, ~3\pi/8$ and $\pi/2$ respectively.  One has $\kappa_{1,2} \in [-1.5,~1.5]$ and 
$\kappa_1$, $\kappa_2$  can not  both take small values to give rise to a correct dark matter relic density.  
By assuming $\kappa_1 =\kappa_2$ and  degenerate mediator masses, we show in Fig. \ref{relicPlots} (b) contours 
of the dark matter relic density,  with  the red, yellow, brown, blue, magenta, cyan and  orange colored contours 
corresponding to $m_\psi =5 ~{\rm GeV},~10~{\rm GeV}, ~20~{\rm GeV}, ~50~{\rm GeV}, ~100~{\rm GeV}, ~200~{\rm GeV} $ 
and $500$ GeV respectively. It shows that  the heavier the dark matter is,  the larger the annihilation cross section 
will be,  such that  larger mediator masses or smaller couplings will be required to get a correct relic density.  
This can also be seen from  Fig. \ref{relicPlots} (c)  and (d), where we show the correlation between the dark matter
mass and the charged mediator masses (Fig. \ref{relicPlots} (c))  as well as  the correlation between the neutral 
mediator masses and charged mediator masses (Fig. \ref{relicPlots} (d)).  For the input of  other parameters of 
Fig. \ref{relicPlots} (c)  and (d) see the caption  for details.

\begin{figure}
  \centering
  \includegraphics[width=0.5\textwidth]{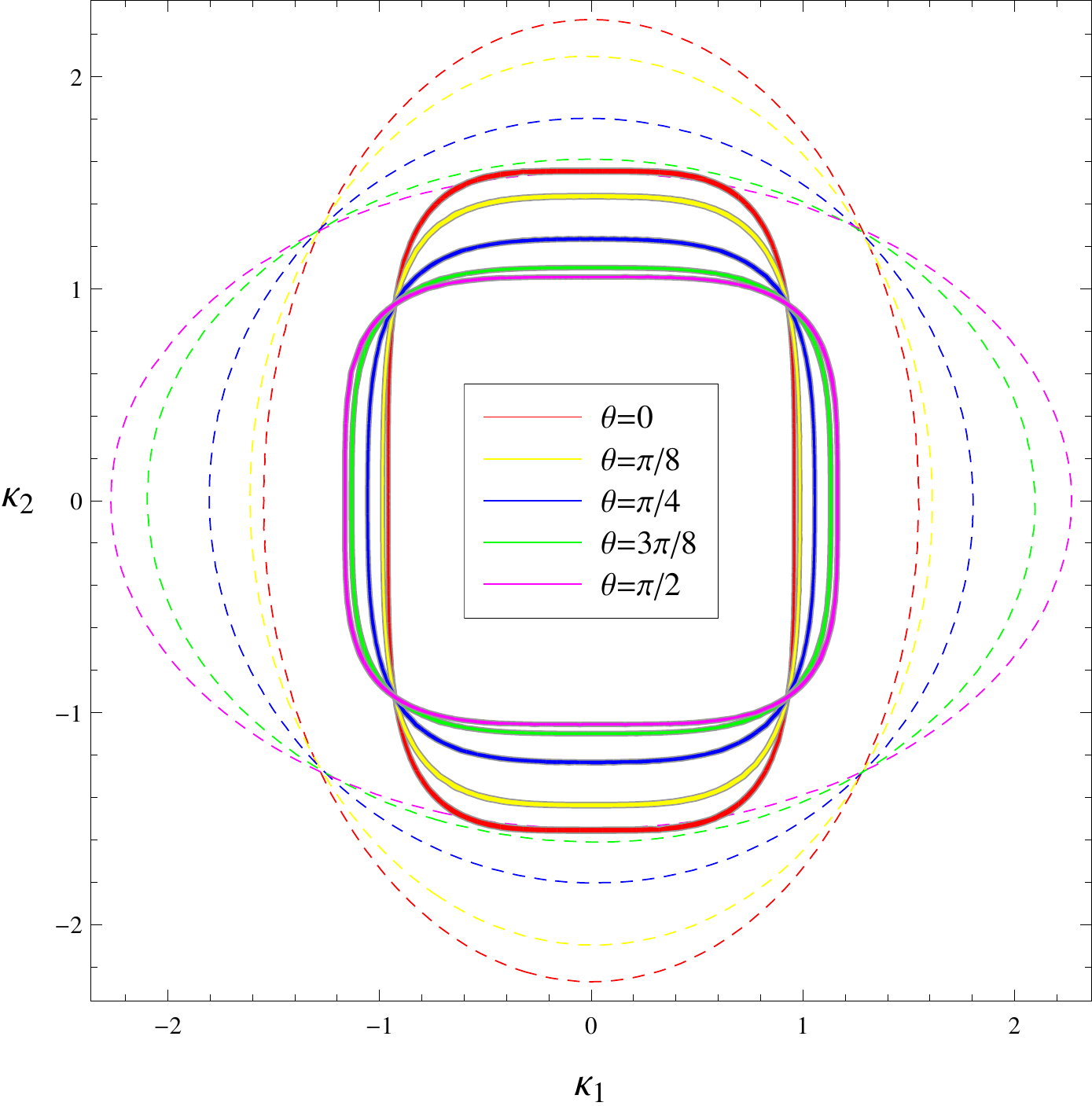}
  \caption{
LUX limit as a dashed line for each solid relic density contour in the $\kappa_1$, $\kappa_2$ plane for different $\theta$. Relic density contours are within two standard deviations of the Planck measured central value and regions outside the dashed lines are excluded at 95\% C.L. by the LUX. The other parameters are fixed to be $m_{\psi}=100$ $\text{GeV}$, $\hat{m}_{1}=400$ $\text{GeV}$, $\hat{m}_{2} =600$ $\text{GeV}$, $m_{\rho}=m_{\eta}=700$ $\text{GeV}$ and $\lambda_2=\lambda_5=0$.
\label{k_with_dd_limit}
}
\end{figure}

\subsection{Direct detection}
Notice that flavored dark matter models may help to release the tension between the observed dark matter relic density and 
constraints from direct detections, which detect dark matter scattering from nuclei in underground laboratories. In our model, dark matter couples 
to nucleons through loop induced electromagnetic form factors of the dark matter as well as loop induced dark 
matter-Higgs interactions.  The effective interactions of the dark matter with nucleon take the following form~\cite{Chang:2014tea,Hamze:2014wca,:2015fqa}
\begin{eqnarray}
f_r\bar \psi \gamma^\mu \psi \bar N \gamma_\mu N +f_h \bar \psi \psi \bar N N + f_m^1\bar \psi i\sigma^{\mu\nu} \psi { q_\nu \over q^2 } \bar N K_\mu N +f_m^2 \bar \psi i\sigma^{\alpha\mu} \psi { q_\alpha q_\beta \over q^2 } \bar N i\sigma^{\beta \mu} N \label{effectnucl}
\end{eqnarray}
with $q^{\mu}$ being the momentum transfer from nucleon to dark matter and $K^{\mu}$ defined as the summation of momenta of incoming and outgoing nucleon. 
The Wilson coefficients are given by 
\begin{eqnarray}
  f_r^N = e Q_N b_\psi \; , \hspace{0.5cm} f_h^N = f_\psi^h {m_N \over m_h^2 v}\left( \sum_{q=u,d,s}  f_{T_q}^N + {2\over 9} f_{TG}^N\right)\; ,  \hspace{0.5cm}  f_m^1 = \frac{e Q_N \mu_\psi}{2m_N}\; ,  \hspace{0.5cm} f_m^2 = -\frac{e \tilde{\mu}_N \mu_\psi}{2m_N} ,\nonumber 
\end{eqnarray}
where $Q_N$ is the charge of the nucleon, $\mu_\psi$ and $b_\psi$ are the magnetic moment and charge radius of the dark matter 
respectively, $\tilde \mu_N $ is the nucleon magnetic moment, that is, $\tilde{\mu}_p \approx 2.80$ and $\tilde{\mu}_n \approx -1.91$. 
Finally $f_{\psi}^h$ is the effective dark matter-Higgs coupling,
\begin{eqnarray}
  f_{\psi}^h = \sum_{ij=1}^2  {c_{ij} m_\psi \over 32 \pi^2} \int_0^1 dx  \int_0^{1-x} dy {  1-x \over (1-x-y)\hat m_{i}^2 + y\hat m_j^2 + (x^2-x) m_\psi^2 } ,
  \label{HiggsWC}
\end{eqnarray} 
where $
c_{11} \approx \zeta_{1}  v (\lambda_2 c^2_\theta + \lambda_5 s^2_\theta{+}2\Lambda s_{\theta}c_{\theta}/v)$, $ c_{22} \approx \zeta_2  v[\lambda_2 s^2_{\theta} + \lambda_5 c_\theta^2 {-}2\Lambda s_{\theta}c_{\theta}/v) ]$, 
$c_{12} =c_{21} = s_{\theta} c_{\theta} (\kappa_1^2-\kappa_2^2) [ v s_{\theta} c_{\theta} (\lambda_5-\lambda_2 ) + \Lambda c_{2\theta}]$  and $x$, $y$ are Feynman parameters.  We have neglected the  $Z$ mediated interactions in Eq.(\ref{effectnucl}) since they are subdominant compared with photon mediated processes. 

\begin{figure}
  \centering
\scalebox{0.95}{
\begin{subfigure}[b]{0.49\textwidth}
  \includegraphics[width=\textwidth]{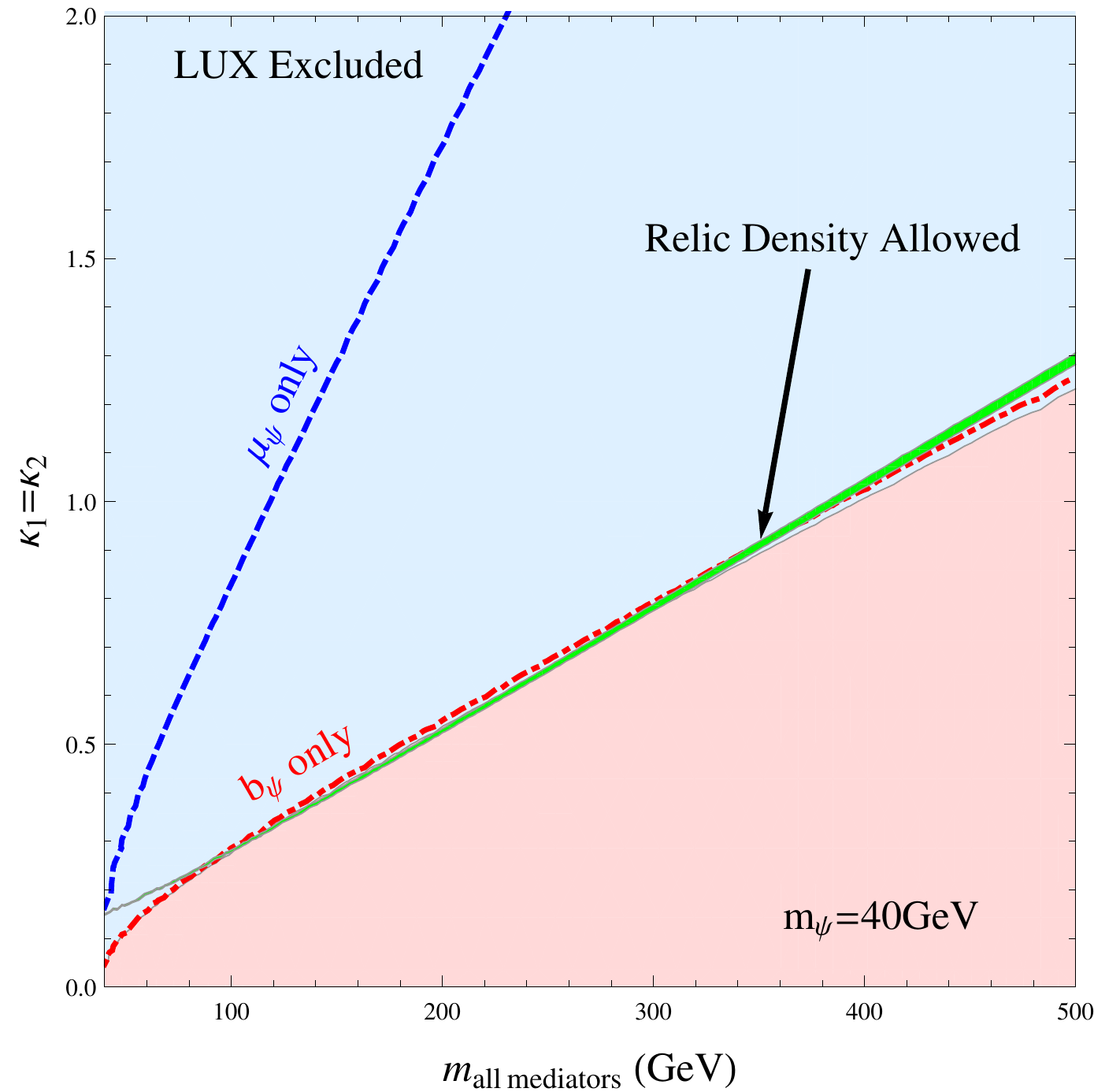}
  \caption{}
\end{subfigure}
}
\scalebox{0.95}{
\begin{subfigure}[b]{0.49\textwidth}
  \includegraphics[width=\textwidth]{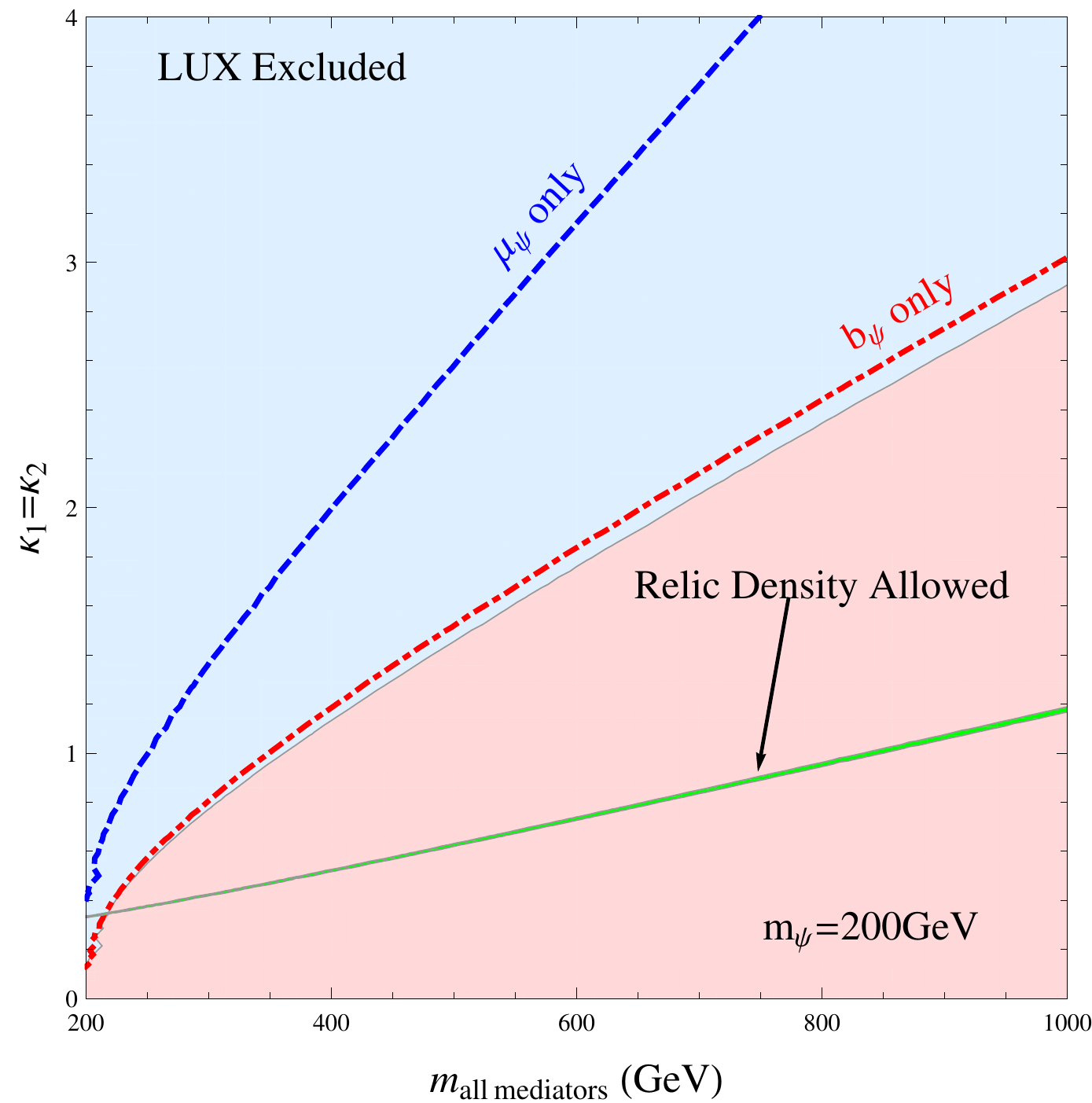}
  \caption{}
\end{subfigure}
} \\
\scalebox{0.95}{
\begin{subfigure}[b]{0.49\textwidth}
  \includegraphics[width=\textwidth]{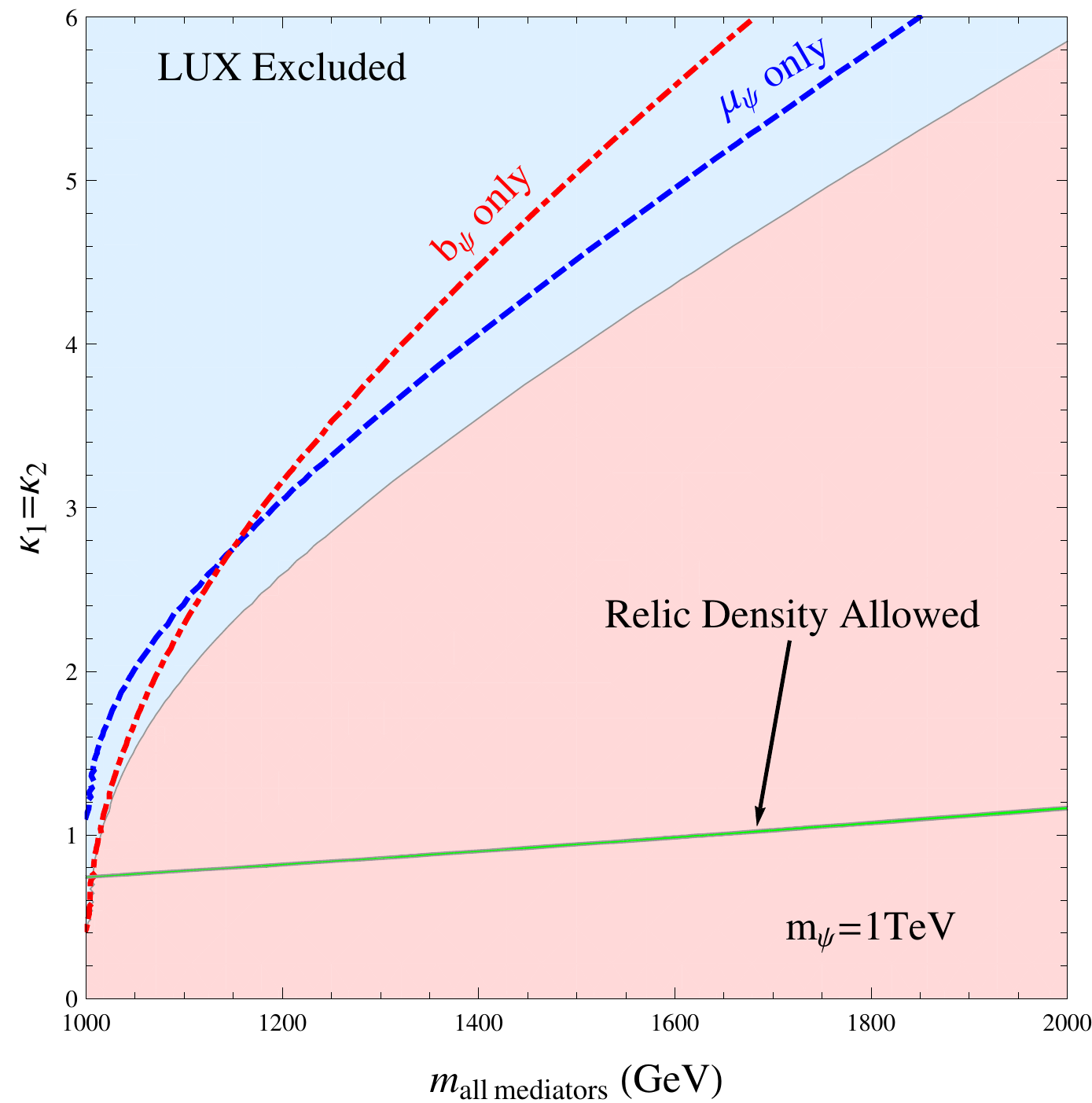}
  \caption{}
\end{subfigure}
}
\scalebox{0.95}{
\begin{subfigure}[b]{0.49\textwidth}
  \includegraphics[width=\textwidth]{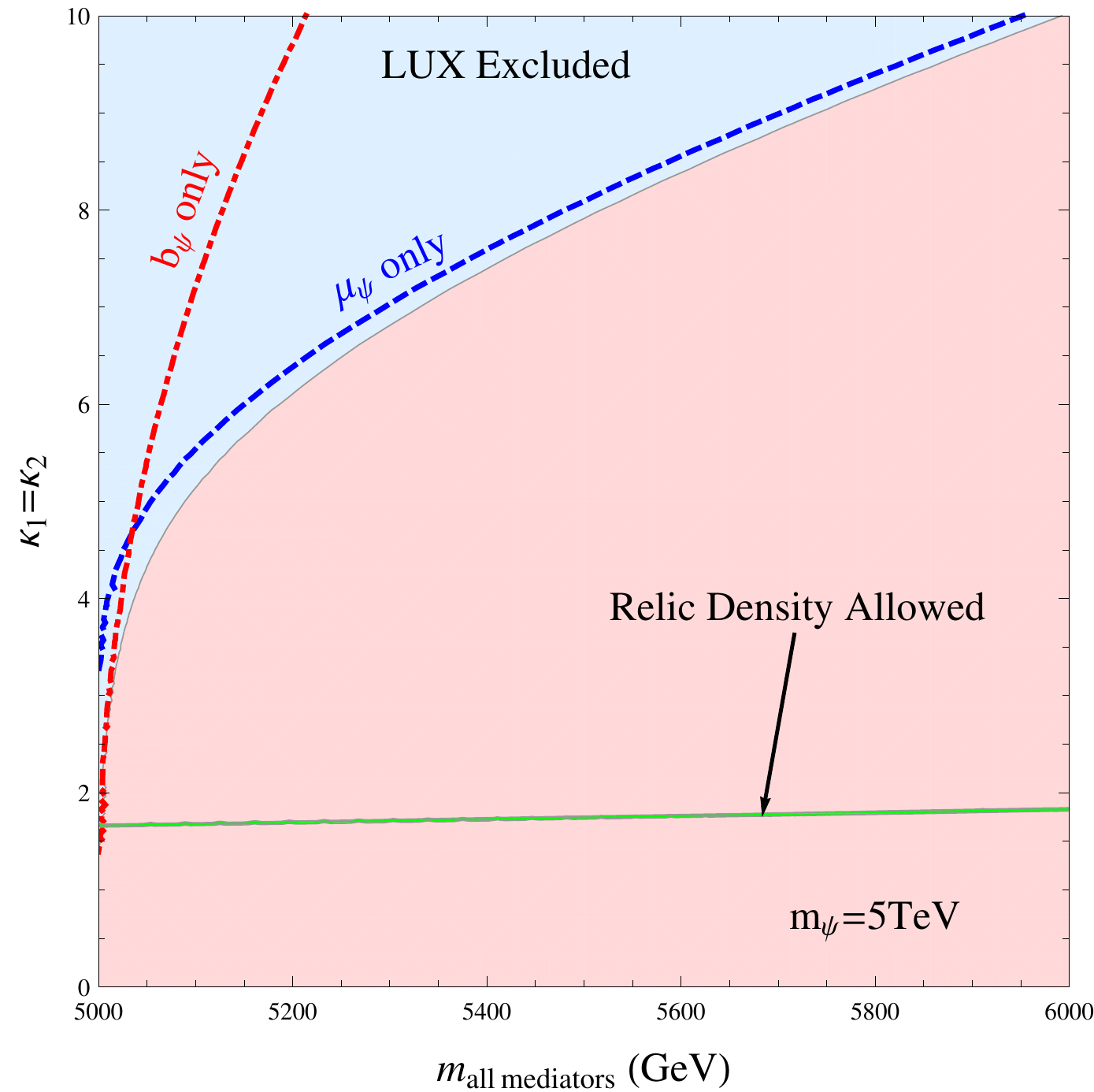}
  \caption{}
\end{subfigure}
}
\caption{\label{kvsMassDD} 
LUX excluded and relic density allowed regions for dark matter with different masses in the couplings versus mediator 
mass plane. LUX excluded regions at 95\%C.L. are shown in light blue while light red regions are allowed. The ``$\mu_{\psi}$ only'' 
dashed line is the LUX limit retaining only the contribution of magnetic dipole moment while ``$b_{\psi}$ only'' corresponds to including
 only charge radius contribution. The green contours are relic density allowed regions within two standard deviations of the Planck 
central value. In all plots we assume $\kappa_1 = \kappa_2$, equal mediator masses and $\lambda_2=\lambda_5=0$.
}
\end{figure}

The momentum dependence induced by the magnetic moment term makes it impossible to factorize the differential event rate 
into the product of the elastic cross section and momentum integration. We therefore need to calculate the differential 
rate numerically then translate the cross section into event rate in experiment.  One more complexity arises since the 
operators shown above go beyond the traditional spin-independent and spin-dependent characterization of dark matter 
nucleus scattering and therefore more nuclear responses are involved~\cite{Gresham:2014vja}. The corresponding classification 
of the underlying non-relativistic operators responsible for dark matter nucleon scattering as well as the identification and 
calculation of nuclear responses for finite-sized nucleus have been performed systematically in an effective field theory 
framework in Ref.~\cite{Fitzpatrick:2012ix,Anand:2013yka} following earlier work in Ref.~\cite{Fan:2010gt}. This framework 
has been implemented in the public code~\cite{DelNobile:2013sia} together with statistical analysis for each experiment. 
We therefore use this code in our analysis and refer the reader to the above literatures for details.

{ We add constraints of the dark matter direct detection to the relic density plot in Fig. 3 (a), and the new plot is 
shown in Fig.~\ref{k_with_dd_limit}.  For each relic density contour, we plot its  corresponding limit from the LUX~\cite{Akerib:2015rjg} at the 
95\% C. L., which is shown as a dashed line with the same color as the relic density contour. We can see that the LUX allowed 
maximum magnitude of $\kappa_{1,2}$ is $1\sim 2$ while the corresponding relic density allowed magnitudes are smaller, which are thus allowed by the LUX.
Notice that all direct detection limit lines intersect at four points 
when $\vert \kappa_1 \vert = \vert \kappa_2 \vert$ just like the case of the
relic density contours. This is because $\mu_{\psi}$ and $b_{\psi}$ are both independent on the mixing angle $\theta$ in this 
scenario and the contribution arising from the anapole moment is velocity suppressed and thus 
negligible.

We also show representative plots in Fig.~\ref{kvsMassDD} on the correlations between the coupling strength $\kappa_1=\kappa_2$ 
and the totally degenerate mediator masses, where subfigures (a), (b), (c) and (d) correspond to taking the dark matter mass as 
$40~\text{GeV}$, $200~\text{GeV}$, $1~\text{TeV}$ and $5~\text{TeV}$, respectively.  The other parameters are fixed to be 
$\lambda_2=\lambda_5=0$, and $\Lambda=0$ as a result of the assumed degenerate charged scalars. So the effective dark matter-Higgs
coupling is exactly zero in this scenario.  Generally the Higgs mediated contribution is suppressed by the Higgs mass squared and 
thus subdominant  compared with contributions of electromagnetic form factors~\cite{Ibarra:2015fqa}. In each plot, light blue(red) 
regions are excluded (allowed) by the LUX at the 95\% C. L.; the green contours represent regions where the dark matter relic density 
is consistent with the measured value within $2\sigma$ level;  the red dot-dashed (blue dashed) line is the LUX limit when considering only
the contribution of charge radius (magnetic moment).  One can see from these figures the roles played by the magnetic moment and the charge
radius in the dark matter direct detections. For a relatively light dark matter, the charge radius dominates the contribution to the 
direct detection;  while for the superheavy dark matter, the magnetic moment plays more important role.   This is because the charge 
radius operator is dimension six while the magnetic moment operator is dimension five.   It also shows that the dark matter should be
around $50$GeV or heavier  to release the tension  between the measured dark matter relic density and constraints from the LUX.

\comment{\textcolor{red}{
Of course, the above case is is only a special region in the whole parameter space and more flexibilities can be obtained
by relaxing the assumptions made.
}
}}

\subsection{Higgs Couplings}

\begin{figure}
{
  \includegraphics[width=0.49\textwidth]{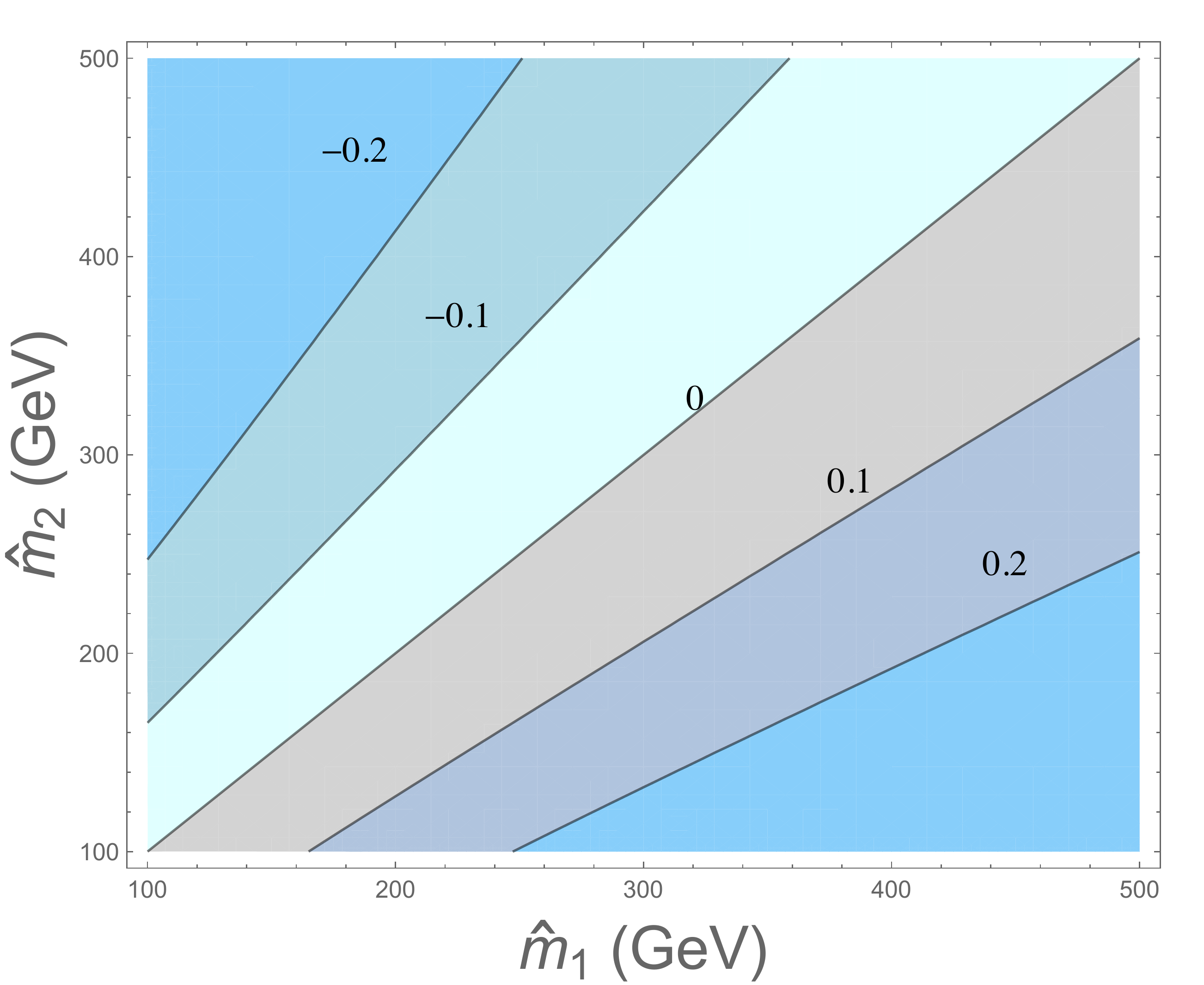}
  \includegraphics[width=0.49\textwidth]{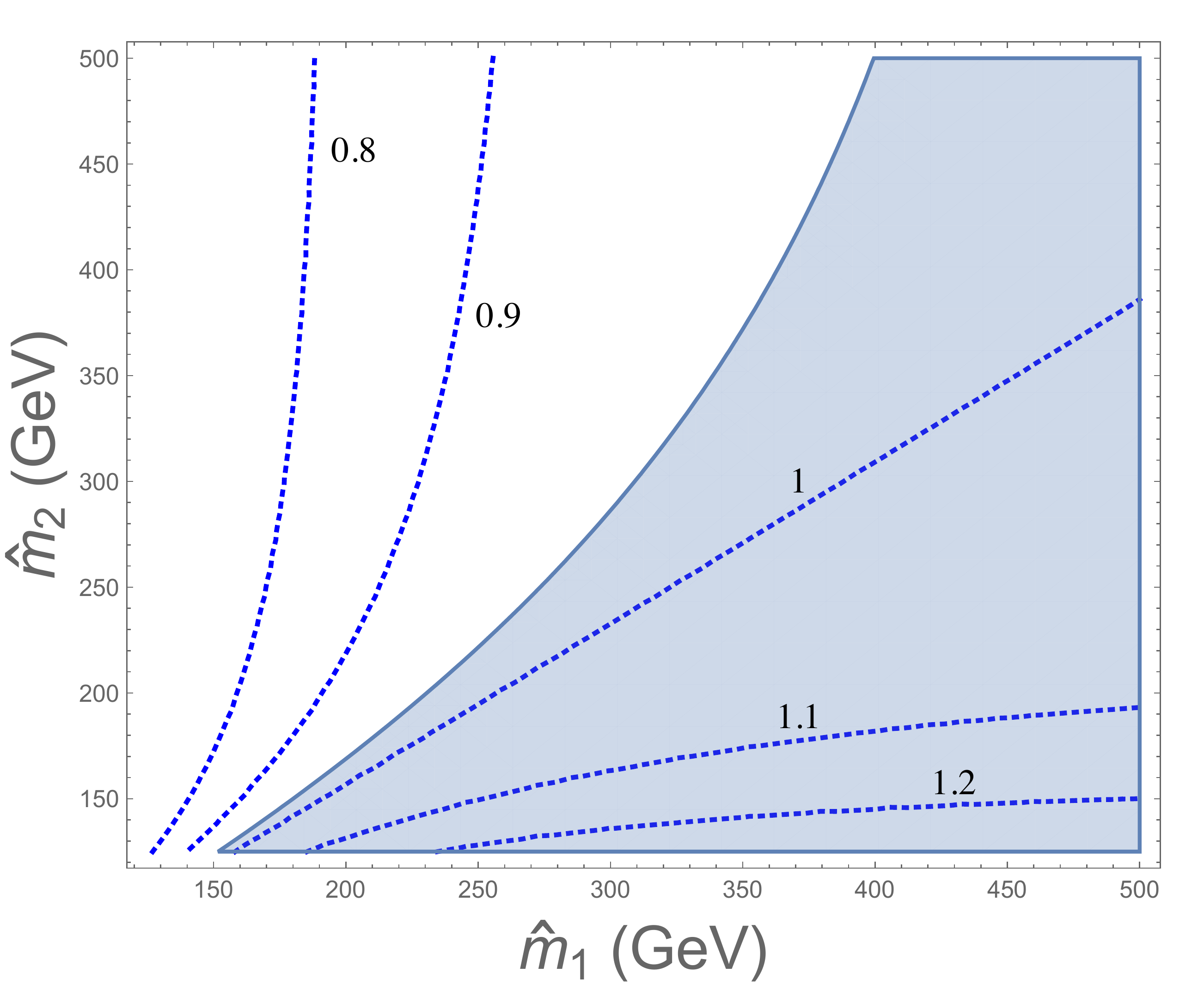}
}
\caption{Left panel:  Contours of $m_\tau^{\rm loop}/m_\tau$ in the $\hat m_1^{} -\hat m_2^{}$ plane;  Right panel:  Contours of  $\mu_{\gamma\gamma}$ in the  $\hat m_1-\hat m_2$ plane, the cyan color marked region satisfy the combined constraint given by the ATLAS and CMS. \label{ratio}
}
\end{figure}

Precision measurement of the Higgs couplings is one of the most important tasks in the future Higgs factory. 
The tau lepton Yukawa coupling was measured by the ATLAS and CMS collaborations, 
whose results are not so consistent with the SM prediction: $\mu_{\tau\tau}=1.4\pm0.4$ by the ATLAS collaboration~\cite{Aad:2015vsa}
and $0.78\pm0.27$ by the CMS collaboration~\cite{Chatrchyan:2014nva}.  In our model, the tau lepton mass arises from two parts: the 
Yukawa interaction induced term, i.e., $m_{Y}^\tau=y_\tau v/\sqrt{2}$, as well as loop corrections,  $m_\tau^{\rm loop}$, 
mediated by the dark matter and two charged scalars.  The mass can  be written as
\begin{eqnarray}
  m_\tau^{} \approx  y_\tau   v/\sqrt{2} + {c_{\theta}s_{\theta}\kappa_1^{} \kappa_2^{} m_\psi^{}  \over 16 \pi^2 } \left[ { \hat m_{1}^2 \over  \hat m_1^2 - m_\psi^2  }\ln\left( {\hat m_1^2 \over m_\psi^2}\right)-{ \hat m_{2}^2 \over \hat m_{2}^2 - m_\psi^2  }\ln\left( { \hat m_{2}^2 \over m_{\psi}^2}\right)\right] \; ,
\end{eqnarray}
where we have neglected terms proportional to $m_Y^\tau$ in the calculation of $m_\tau^{\rm loop}$. We show in the 
left panel of Fig. \ref{ratio}  contours of $m_\tau^{\rm loop}/m_\tau$ in the $\hat m_1-\hat m_2 $ plane by setting
$\kappa_1 =\kappa_2 =1 $, $c_\theta =0.6$ and $m_\psi=100~{\rm GeV}$ which are consistent with dark matter constraints.
It is clear that $m_\tau^{\rm loop}$ can be ${\cal O} ( 10\%)$ of the total tau mass. 

\begin{figure}
{
  \includegraphics[width=0.49\textwidth]{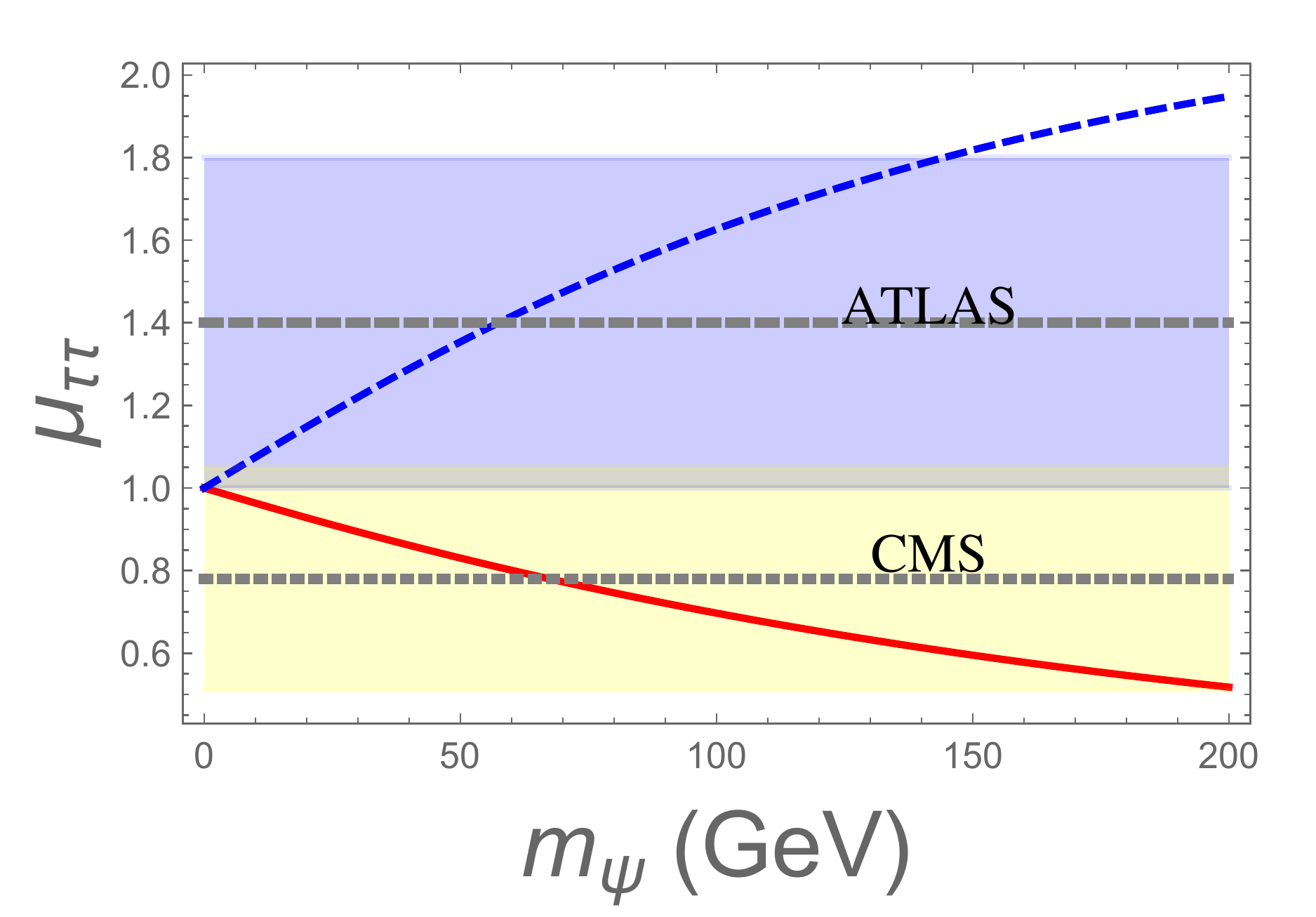}
}
\caption{Signal rate of Higgs to tau tau relative to the SM expectation as a function of the dark matter mass. We set  $c_\theta=0.5$, $\lambda_2=\lambda_5=0.1$,  
$\hat m_1 =400~{\rm GeV}$ and $\hat m_2 =600~{\rm GeV}$ as well as $\kappa_1^{} =-\kappa_2^{} =1$ for the red solid curve 
and  $\kappa_1^{} =\kappa_2^{}=1$ for the blue dashed curve.   \label{higgstautau}
}
\end{figure}

The branching ratio for the Higgs decaying into $\tau \bar \tau$ can be approximately written  as
\begin{eqnarray}
{\rm BR} (h\to \tau \tau) \approx {m_h \over 16\pi \Gamma_{\rm tot}} \left| y_\tau + \sqrt{2 } \xi_\tau\right|^2
\end{eqnarray}
where $m_h$ is the SM Higgs mass, $\Gamma_{\rm tot} =4.1\times 10^{-3}~{\rm GeV}$ is the SM Higgs decay width
 and  the loop induced coupling can be written as
\begin{eqnarray}
{\xi}_\tau=\sum_{ij=1}^2 {y_{ij}  m_\psi \over 16 \pi^2 } \int_0^1 dx \int_0^{1-x} dz {1 \over  xm_\psi^2 + z \hat m_i^2 + (1-x-z)\hat m_j^2 -z(1-x-z) m_h^2 } ,
\end{eqnarray}
with $y_{11} = \kappa_1 \kappa_2 c_\theta s_\theta [( \lambda_2 c_\theta^2 + \lambda_5 s_\theta^2 ) v+\Lambda s_{2\theta}]$, $y_{22} = -\kappa_1 \kappa_2 c_\theta s_\theta [( \lambda_2 s_\theta^2 + \lambda_5 c_\theta^2 ) v-\Lambda s_{2\theta}]$ and $y_{12}=y_{21}= 1/2\kappa_1 \kappa_2 c_{2\theta} [s_{\theta}c_{\theta}v (\lambda_5-\lambda_2) + \Lambda c_{2\theta}]$.   
We plot in Fig.~\ref{higgstautau} the signal rate $\mu_{\tau \tau}$ associated with Higgs measurements, relative to the 
SM Higgs expectation, as a function of the dark matter mass by setting $c_\theta=0.5$, $\lambda_2=\lambda_5=0.1$,  
$\hat m_1 =400~{\rm GeV}$ and $\hat m_2 =600~{\rm GeV}$ as well as $\kappa_1^{} =-\kappa_2^{} =1$ for the red solid curve 
and  $\kappa_1^{} =\kappa_2^{}=1$ for the blue dashed curve. The dashed and dotted horizontal lines represent central values 
given by the ATLAS and the CMS respectively with light blue and light yellow bands corresponding to uncertainties at the $1\sigma$ level.  
It should be mentioned that  $\mu_{\tau\tau}^{}$ can be significantly changed for some extreme scenarios and the modification can also be tiny for other cases (small  $\kappa_{1,2}$, light dark matter and heavy degenerate charged scalars).

Due to the existence of charged scalars, the Higgs to diphoton decay width is slightly modified.  The decay rate can be 
written in terms of couplings of the SM Higgs with new charged  scalars:
\begin{eqnarray}
\Gamma (h\to \gamma \gamma ) = {G_F \alpha^2 m_h^3 \over 128 \sqrt{2} \pi^3 } \left| -6.48+ \sum_{i=1}^2 { v  c_{ii} \over 2 \zeta_i\hat m_i^2    } A_0 \left(  4 \hat m_i^2 \over m_h^2 \right)\right|^2 ,
\end{eqnarray}
where $-6.48$ is the contribution of the $W$ and top loops and the second term is the contribution of two new charged scalars 
with the definition of the loop integral function $A_0 (x)$ following conventions of Ref~\cite{Carena:2012xa} 
\begin{eqnarray}
A_0(x) = -x^2 \left [   {1\over x} -f({ x^{-1}} )  \right] \hspace{0.3cm}  {\rm with} \hspace{0.3cm}    f(x) \equiv
  \left\{
    \begin{array}{ll}
      \arcsin^2(\sqrt{x}),  &   \quad \text{for} \quad  x > 1,  \\
      -\frac{1}{4} \left(\ln\frac{1+\sqrt{1-x^{-1}}}{1-\sqrt{1-x^{-1}}}-i\pi \right)^2,  
      &  \quad \text{for} \quad  x <  1.
    \end{array}
  \right. 
\end{eqnarray}


We plot in the right panel of Fig.~\ref{ratio} contours of  $\mu_{\gamma\gamma}$ in the $\hat m_1 -\hat m_2$ plane by
setting $\lambda_2=\lambda_5=0.5$ and $c_\theta =0.8$.  The green dashed lines from the left 
to the right correspond to $\mu_{\gamma \gamma} =0.8,~0.9,~1.0, ~1.1, ~1.2$ respectively.  The cyan color marked region 
satisfies the current combined bound given by the ATLAS and CMS collaborations, $\mu_{\gamma\gamma}^{} =1.15\pm0.18$, where  
$\mu_{\gamma \gamma} =1.17\pm0.27$ by the ATLAS~\cite{Aad:2014eha} and  $\mu_{\gamma\gamma}=1.14_{-0.23}^{+0.26}$ by
the CMS \cite{Khachatryan:2014ira}.  It should be mentioned that the future improved measurements of $\mu_{\gamma \gamma}$
may put more severe constraint on  couplings of the Higgs to new charged scalars.
  
Finally, lets comment on the collider searches of this model. The collider signals of lepton portal dark matter models are
events with charged lepton pairs and missing energy. It was showed in Ref.~\cite{Agrawal:2011ze} that these models have clear signals 
above the SM background in certain parameter space at the LHC.  Searches for signatures of our model at the LHC and lepton colliders such 
as CEPC or ILC, which are interesting but beyond the reach of this paper, will be shown in a future study. 

\section{\label{sec:conclude}Concluding remarks}

Lepton-flavored dark matter is interesting and appealing for many aspects. In this paper we focused 
on the phenomenology of the tau-flavored Dirac dark matter model.  The electromagnetic form factors 
of the dark matter which are crucial for the dark matter direct detections, were calculated  in the 
case  where there are two types of dark matter - third generation lepton Yukawa 
interactions. Our study shows that the tension between the observed dark matter relic density and constraints of dark matter
direct detections are highly loosed. The charge radius dominates the contributions to the dark 
matter direct detection for the light dark matter case, while the magnetic moment plays more important role
for heavy dark matter case.  In addition the tau Yukawa coupling  can be significantly changed 
in this model, and the one-loop induced  tau mass can be  ${\cal  O} (10\%) $ of the total mass.  As a result, the 
signal rate of $h \to \bar \tau\tau$ , relative to the SM expectation, measured by the LHC,  can be explained 
in this model.  The Higgs to diphoton ratio is also slightly changed but is still consistent with the 
current LHC constraint.


\begin{acknowledgments}
The authors are indebted to  Prof. Michael J. Ramsey-Musolf for helpful discusssions and encouragements. 
This work was supported in part by DOE Grant DE-SC0011095.  Huai-Ke Guo was also supported by the China Scholarship
Council.  We thank Chien Yeah Seng  for reading the manuscript.

\end{acknowledgments}

\end{document}